\documentclass[12pt,preprint]{aastex}

\shorttitle{A Selection of Giant Radio Sources from NVSS}
\shortauthors{Proctor}

\newcommand{\noprint}[1]{}

\begin{document}


\title{A Selection of Giant Radio Sources from NVSS}

    
\author{D. D. Proctor}
\affil{Visiting Scientist, Lawrence Livermore National Laboratory, L-405,\\
7000 East Avenue, Livermore, CA, 94550; proctor1@llnl.gov}
    
    

\begin{abstract}
Results of the application of pattern recognition techniques to the problem of identifying Giant Radio Sources (GRS) from the data in the NVSS catalog are presented and issues affecting the process are explored.
Decision-tree pattern recognition software was applied to training set source pairs developed from known NVSS large angular size radio galaxies.
The full training set consisted of 51,195 source pairs, 48 of which were known GRS for which each lobe was primarily represented by a single catalog component.   The source pairs had a maximum separation of $20\arcmin$ and a minimum component area of 1.87 square arc minutes at the 1.4 mJy level.  
The importance of comparing resulting probability distributions of the training and application sets for cases of unknown class ratio is demonstrated.
The probability of correctly ranking a randomly selected (GRS, non-GRS) pair from the best of the tested classifiers was determined to be 97.8\underline{+}1.5\%.
The best classifiers were applied to the over 870,000 candidate pairs from the entire catalog. 
Images of higher ranked sources were visually screened and a table of over sixteen hundred candidates, including morphological annotation, is presented.  These systems include doubles and triples, Wide-Angle Tail (WAT) and Narrow-Angle Tail (NAT), S- or Z-shaped systems, and core-jets and resolved cores.  While some resolved lobe systems are recovered with this technique, generally it is expected that such systems would require a different approach.

\end{abstract}


\keywords{astronomical data bases: miscellaneous --- astronomical data bases: catalogs --- galaxies: general --- methods: data analysis --- methods: statistical --- techniques: image processing}



\section{INTRODUCTION}

Using the Karl G. Jansky Very Large Array (VLA), the NRAO VLA Sky Survey (NVSS) \citep{Condon98} is a 1.4 GHz continuum survey covering the entire sky north of -$40\degr$ declination with a resolution of $45\arcsec$ FWHM.  The associated catalog of discrete sources from this survey contains over 1.8 million entries.
This catalog was generated by fitting flux densities of all significant peaks with elliptical Gaussians.  Thus it is a catalog of components, where a single physical system may have multiple catalog entries.  

While the NVSS survey and catalog have been out since 1998, few large scale attempts to identify GRS in the data have been reported.  
\citet{Lara01} examined NVSS images above $+60\arcdeg$ (0.842 steradians or approximately $8\%$ of the NVSS survey area) and presented a sample of 84 large angular size radio galaxies.  This suggests that approximately one thousand similar systems should be present in the entire NVSS survey area.
\citet{Machalski01} compiled a list of 40 candidates of \citet{Fanaroff-Riley74} type II (FRII) or FRI/II with 1.4 GHz flux density on the NVSS maps of $S_{1.4} < 500$ mJy and sky area of 0.47 steradians.
A recent attempt by \citet{Solovyov11} to develop an algorithm for selection of large, faint sources from the catalog was limited to sources between $4\arcmin$ and $6\arcmin$ separation, major axis of source lobes greater than $1\arcmin$, and integrated flux density $\lesssim$ 100 mJy for at least one of the components.  
They reported 61 candidates for the initial stage of their work and subsequently reported on radio and optical identifications of 50 Giant Radio Galaxies (GRG) from that candidate list \citep{Solovyov14}. 
\citet{Andernach12} has made a first visual inspection of approximately 3500 images of NVSS, SUMSS and WENSS radio surveys to search for such sources.  Those results are pending.

For this report, GRS will be interpreted broadly to include any physically associated radio system with projected angular size $\geq 4\arcmin$.  Thus GRS may include star-forming regions of, presumably nearer, galaxies as well as the other usual types.  GRG are typically defined as having an overall projected size of $\geq$ 1 Mpc, thus identification of a GRS as GRG depends upon a distance determination, a determination beyond the scope of this paper.

\citet{Lara01} confirmed 37 of the 84 GRS from their sample to be GRG.
GRG are of interest in the study of evolution of galaxies, jet interaction with intergalactic medium, and testing consistency with the unified scheme for radio galaxies and quasars, as well as verifying cosmological hydrodynamical simulations.  See for example, \citet{Ishwara99} and \citet{Malarecki13} for more details.  


The task of identifying GRS from the catalog data will be considered and the feasibility of the pattern recognition process to significantly expand the number of GRS and GRS candidates will be explored.
Reported here are results of an initial stage of work to apply pattern recognition techniques to the problem of identifying large radio galaxy systems for which each lobe is primarily represented by a single component in the NVSS catalog.
GRS with multi-component lobes are only found serendipitously with this procedure.   Finding systems with multi-component lobes could be attempted with procedures similar to that used for finding groups in FIRST \citep{Proctor11} or the morphological-operator approach employed by \citet{Santiago-Bautista13}.  It is expected that multiple approaches will be necessary to fully mine the information in the database.

The remainder of the paper is organized as follows.
Section 2 contains general comments and background on pattern recognition procedures, including training set construction, feature set selection, decision-tree classifiers, cross validation process and classifier evaluation. 
Section 3 discusses training set results for variations in the training set, feature set, and classifier options.  Section 4 discusses results of applying the best classifier ensembles found to the pairlist from the entire catalog, shows consistency between certain training and application sets, and gives a table and mosaic of resulting candidates.  Finally, discussion and summary are presented in Section 5.  

It should be noted the intensity scaling of the figures shown in this paper have often been chosen to enhance features of interest.  Unless otherwise indicated, the cutouts shown in the mosaics are $20\arcmin$ in size.  
Note that only the tabulated coordinates that are associated with an NVSS catalog source have the
significant figures of the NVSS catalog itself.  The remainder are given to the same number of digits as the NVSS catalog for tabular consistency and convenience.
The key to the annotations used in the candidate table is given in the appendix.


\section{PATTERN RECOGNITION PROCEDURES}

This section presents an overview of pattern recognition procedures.
The pattern recognition process involves four steps:  creation of a training set, selection of features, selection and training of the classifier and evaluation of the classifier.  For those unfamiliar with the process, these steps are discussed in detail in the remainder of this section. 

\subsection{\it Training set construction}

 For this study, a training set was constructed from the population of NVSS catalog sources above $+60\arcdeg$ declination since that was what was available for a verifiable training set.  The GRS target class was developed from
the sample of 84 large angular size radio galaxies presented by \citet{Lara01}.   In constructing their sample, they visually inspected all NVSS maps above $+60\arcdeg$ declination and pre-selected 122 map features apparently related to a single physical source, for which the total flux density was $\ge 100$ mJy and the angular extension was larger than 4\arcmin, as measured along the 'spine' of the source.  They then observed these sources with the VLA to confirm large angular size radio galaxies and reject those objects which were result of superposition of multiple adjacent sources.  This resulted in 84 confirmed GRS and generally consisted of galaxies of angular size $\geq 4\arcmin$ and total flux density greater than 100 mJy at 1.4 GHz.  It is noted that although a few sources fulfilling the requirements of the sample may have been missed due the the existence of "holes" in the maps available at the time, the list was felt to be sufficiently complete to demonstrate proof of principle for the pattern recognition process.
From the \citet{Lara01} list of 84 sources, 48 were selected for the pattern recognition training set.  These 48 sources were those for which each lobe was primarily represented by single catalog component.  The fitted version of the catalog was used both to avoid deconvolution issues and to have a direct correspondence to the images.  Comparison with deconvolved parameters is left for future examination.
Using the catalog parameters, the model silhouette size at the 1.4 mJy level was determined for each GRS component. 
The silhouette size is defined for this study as the area enclosed by the contour of the fitted model at the 1.4 mJy level. 
The minimum size noted for the larger-area component of each pair was 2.43 square arc minutes and minimum size for the smaller component was 1.87 square arc minutes. 
Components of the remaining resolved-lobe GRS were excluded from the training set. 
Using the minimum silhouette sizes found for the Lara GRS sample, a list of source pairs for the training set was constructed such that the first member of the pair was at least 2.43 square arc minutes and the second member was at least 1.87 square arc minutes, with the second member of the pair within $20\arcmin$ of the first and less than or equal to first in area.
Aside from components of the resolved GRS, every source pair that satisfied the stated conditions were included in the training set.
This resulted in a list of 51195 pairs for the training set, of which 48 were the GRS selected from \citet{Lara01} sample.

Normally it is desired to construct a training set from a random sample from the entire population.  This ensures that the statistics generated for the training set will be extendable to the remaining population (the application set) and that the resulting classifier produces results representative of the population.  As the case here, this is not always feasible and subsequent evaluation of results, as discussed below, is required.  We note that there is considerable discussion in the literature concerning the effect of skew (the class ratio, non-target to target), on classifier performance.   For cases where one class is heavily outnumbered by the the other, as the case for this study, where it is over a thousand-to-one, the classifier may have problems learning the minority class.  Several methods have been proposed to counter that effect, including over-sampling of the minority class and under-sampling of the majority class \citep{Chawla02, Batista04, Chawla08, Cieslak08}.  This topic will be explored and examples of under-sampling the majority class will be presented and discussed.
Comparisons of the resulting classifications for the training set and application set (discussed in Section 4, below) should be made to verify that the results are self-consistent.

\subsection{\it Feature Set Selection} 
The NVSS catalog provides position, peak flux, major and minor axes size, and position angle of the elliptic Gaussian fit for each source.  Preliminary decision-tree training runs used five features - the silhouette size and peak flux of each pair member and the source pair separation.  (It was noted in \citet{Proctor06} that the OC1 classifier was well able to adapt to different functional forms for variables in the feature set assuming all relevant information was available.) Given the relatively small number of GRS in the training set, it was decided to use silhouette size and peak flux for the components, contributing only four features to the feature set versus using the semi-major and semi-minor axes and peak flux, which would require six features, in addition to the pair separation.
The number of source neighbors within $20\arcmin$ of the larger source, with silhouette size greater than 1.871 square arc minutes, was added for later tests as an attempt to include the source environment in the feature set.  
Table~\ref{GRS_features} is a list of features chosen.  
Most of the results reported here used this set of six features, the exception being an eight feature set discussed briefly in Section 3.7.  Larger feature sets may be explored further in future work.  

\subsection{\it Classifier Selection:  Background, Overview, and Algorithm Options}
This section describes background relating to the choice of classifier and gives an overview of the chosen classifier with its options.
For this work, the Oblique Classifier One (OC1) decision tree software of \citet{Murthy94} was chosen.  OC1 has been used for a number of astronomical applications, including cosmic ray identification, star-galaxy separation, quasar candidate selection, X-ray source classification and sidelobe flagging in radio surveys.  See \citet{White08} for a review.  In that reference oblique decision trees were found to represent a good compromise between the demands of computational efficiency, classification accuracy, and analytical value of the results.  It has also been used by the author for sorting triples in the FIRST database, \citep{Proctor03} and comparing pattern recognition feature sets \citep{Proctor06}.

\subsubsection{Classifier Overview}
In general, a classifier is an algorithm that uses a set of numerical features describing members of a population, to separate those members into different classes.
Decision trees are a supervised pattern recognition technique in which statistical methods are applied to typical data of known classification (the training set) to generate rules that can then be applied to the feature values of of unknown objects (the application or test set) to determine their classification.
A decision tree is a data structure that contains tests at branches and classifications at leaf nodes. Typically
a binary tree is constructed in which the test at a branch compares a linear combination of the feature values to zero.  If the test is true, a sample is sent down one branch, if false the other branch.  The process continues until a leaf node is reached that contains a class label.  The job of the algorithm is to use the training set to determine the coefficients of the features in the linear combination at each branch.
Geometrically, this results in the feature space being partitioned by a set of hyperplanes into separate volumes, each with the appropriate label.  OC1, while not the first oblique decision tree algorithm, is freely available\footnote{See http://ccb.jhu.edu/software/oc1/oc1.tar.gz}, open source, and has been found easy to use.  See \citet{Murthy94} for more detail.  While more complex to construct, in general oblique trees are expected to provide a more compact representation.

Application of the resulting classifier to a sample member of unknown classification, results in that member being assigned a probability of being a particular class. 
To determine the probability estimate at a particular leaf node for the positive class, the so-called Laplace estimate has been adopted.  If an object ends at a node with $N_{node}$ training set objects, of which $N_{pos}$ are of the positive class, then the tree's estimated probability in favor of the positive class for the object is ($N_{pos}+1)/(N_{node}+2)$. 
This form is applicable both for pure nodes, where $N_{node} = N_{pos}$ or $N_{node} = N_{neg}$ and nodes of mixed class resulting from any pruning of the tree.
 This was the form adopted by \citet{White00} and \cite{Provost03} and considered by the latter to be the de facto standard, though others still use maximum-liklihood estimates.  \cite{Provost03} found that, while not surprising statistically, the uniformity and magnitude of the improvement using this estimate to be remarkable.

 Five fold cross-validation was adopted.  Cross-validation is a standard pattern recognition technique used to avoid bias that would occur if points used in testing were the same as those used in training.  For five fold cross validation the training set members are divided into five folds (groups) and training set members from four folds are used to generate classifiers for the remaining fold, each fold being classified in succession.  

 If multiple classifiers are generated using some randomization procedure, as in the search for the best hyperplane, the resulting probability estimate is expected to be of improved accuracy and its standard deviation can be calculated.  Such improvement was demonstrated by \citet{Provost03}.
For the current work, ten decision trees were generated with random seeds for each fold, resulting in an ensemble of ten decision trees contributing to probability estimates for each fold for the training set and 50 decision trees contributing to the probability estimates for the entire catalog application set members.  
It should be noted that this is a conditional probability, depending on the classifier, training set, and feature set.  
Rankings based on probability of class membership are needed so that cases most likely to belong to the target class can be considered first.
The ensemble average of the resulting probability estimates for a member will be designated the normalized score or vote.  Ordering this vote, say high to low, for the entire sample, results in a sort-ordered distribution, with index 1 to N, where N is the number of members in the sample.  If the index is divided by the number of sample members it becomes the sample-size normalized index.  The plot of normalized score or vote versus the sample-size normalized index is designated the vote curve.  

For each source i, if $P_{i}$(GRS) is the estimate in favor of
    the source being GRS and $P_{i}$(nonGRS) is the estimate against, then
\begin{eqnarray}
              P_{i}(GRS) + P_{i}(nonGRS) = 1.
\end{eqnarray}
Thus,
\begin{eqnarray}
             {\Sigma^N_{i=1}} P_{i}(GRS)  + {\Sigma^N_{i=1}} P_{i}(nonGRS)  = N,
\end{eqnarray}
     where N is the number of points in the sample under consideration.  Normalizing by N gives
\begin{eqnarray}
          [ {\Sigma^N_{i=1}} P_{i}(GRS) + {\Sigma^N_{i=1}} P_{i}(nonGRS) ] /N  =  1.
\end{eqnarray}
The normalized total in favor is the 'area under the curve' of the sort-ordered, sample-size normalized vote plot.  It allows comparison of vote curves for the training set and application set, as will be shown in Section 4.  See \citet{Proctor03} and \citet{Proctor06} for other examples of vote curve comparisons.

\subsubsection{Classifier Options}
OC1 allows, among other options, choice of pruning portion, impurity measure, order of coefficient perturbation, number of jumps attempted to get out of a local minimum, and the number of restarts at each node to test for the best hyperplane.
Each will be discussed in more detail in the following paragraphs.
  The other OC1 program options should not affect solution results.

\paragraph{Pruning portion:}
  OC1 prunes decision trees by default, to avoid the problem of overfitting.  The only pruning method implemented is error complexity pruning using a separate pruning set.  This option specifies that a chosen fraction of the training set will be used exclusively for pruning.  If the pruning portion = 0, no pruning is done.  

\paragraph{Impurity measures:}
  The impurity measure is the metric used to determine the goodness of a hyperplane location. New nodes are added so as to minimize the "impurity" of the training set members at the node. 
Impurity measures Gini Index, Twoing Rule, Max Minority, Sum Minority and Information Gain are distributed with the OC1 software.  Early tests showed performance of Max Minority and Sum Minority for lower skew training sets was quite poor, and they were dropped from further comparisons.  Two other impurity measures were tested, since they were developed specifically to be skew insensitive.
The impurity measure, designated Flach in tables below, was developed by \citet{Flach03} as a skew insensitive splitting criterion, in particular
\begin{eqnarray}
                 1 - \frac{2 tpr \cdot fpr}{tpr+fpr} - \frac{2(1-tpr)(1-fpr)}{1-tpr+1-fpr}
\end{eqnarray}
is optimized in selecting the split at a node.  Here tpr is the ratio of true positive to total positive instances for the trial split and fpr is the ratio of false positive to true negative.
For the impurity measure denoted Hellinger, the form for relative impurity for the Hellinger distance developed by \cite{Cieslak08} as a skew insensitive measure is
\begin{eqnarray}
                \sqrt{(\sqrt{tpr} - \sqrt{fpr})^2 + ( \sqrt{1-tpr} - \sqrt{1-fpr})^2}.
\end{eqnarray}

\paragraph{Order of coefficient perturbation:}
  At each node OC1 adjusts the coefficients of the hyperplane under consideration.  The default order of adjustment is sequential, which means all the coefficients of a hyperplane are perturbed in order to search for the best value to be found for that coefficient.  The alternatives are best first and random.  For best first, the coefficient that provides the greatest improvement in the impurity measure is done first. For random coefficient perturbation, a coefficient is selected at random and the number of times to pick a random coefficient can be specified.

\paragraph{Number of jumps to escape local minimum:}
When the algorithm cannot improve any of the coefficients of a hyperplane deterministically, it is stuck in a local minimum.  It then attempts a number of jumps out of this  minimum by choosing a random direction and sliding the hyperplane in that direction to improve the impurity measure.

\paragraph{Number of restarts:}
The number of restarts is the number of random hyperplanes at each node examined for best position. This includes the best axis parallel split, if desired.

\subsection{\it Classifier Evaluation}
 The confusion matrix, also called the contingency table, is useful for defining evaluation metrics.  For the two class problem under consideration, the confusion matrix becomes as shown in Table~\ref{conf_matrix}, where TP, FP, FN, TN stand for true positive, false positive, false negative, and true negative counts respectively; PP and PN stand for predicted positive and predicted negative counts; and POS and NEG stand for the number of positive and negative class members respectively, with $N_{ts}$ the training set size, $(POS + NEG = N_{ts})$.  Lower case is used for relative frequencies, e.g. $tp=TP/N_{ts}$.  The true positive rate (hit rate) is defined as tpr = TP/POS and the false positive rate (false alarm rate) fpr=FP/NEG.  Accuracy is defined as $Acc = (TP+TN)/N_{ts}$ and error rate as $Err=(FP+FN)/N_{ts}$.  Another important quantity is the class ratio or skew, defined here as c = NEG/POS.  Without loss of generality, the GRS class is considered the positive class for this paper.

Until recently, classification accuracy has been the primary metric used to evaluate classifiers \citep{Demsar06}. 
As class distribution becomes more skewed, evaluation based on accuracy breaks down, as it is biased toward the dominate class.  It has become common to use Receiver Operating Characteristic (ROC) analysis to make more general comparisons \citep{Ferri02}.
  A ROC graph depicts trade offs between tpr and fpr, typically shown as fpr on the x-axis and tpr on the y-axis.  The Area Under the ROC Curve (AUC) gives the classifier performance as a single scalar and has been shown to be equivalent to the probability that a randomly chosen  (positive class member, negative class member) pair will be correctly ranked (\citet{Hanley82}, \citet{Batista04}, and references therein).  An algorithm for generating a ROC curve from a set of ranked examples was given by \citet{Provost01} and was adopted for this work.  It generates a segment that bisects the area that would have resulted from the most optimistic and most pessimistic orderings of examples with the same score or vote.
Note that a specific classifier instantiation is represented by a single point on the ROC curve, thus
only when one classifier is better over entire performance space can it be declared the better classifier.
Example ROC curves will be shown below.  Comparing classifiers using ROC analysis enables comparison of ranking quality across the entire range of possible class thresholds.  
\citet{Hanley82} discuss the three way equivalence between the AUC, the  Wilcoxon-Mann-Whitney non-parametric test statistic ("the Wilcoxon"), and the probability of a correct ranking of, in the case here, a (GRS, non-GRS) pair.
They also developed a method for estimating the standard error of the difference between the areas of two ROC curves for case of applying different treatments to the same data.  They give 
\begin{eqnarray}
    SE(\hat{AUC_1} -\hat{AUC_2}) = \sqrt{SE^2(\hat{AUC_1)} + SE^2(\hat{AUC_2}) - 2 \cdot r \cdot SE(\hat{AUC_1}) \cdot SE(\hat{AUC_2})} \label{sedif_auc}
\end{eqnarray}
where $\hat{AUC_1}$ and $\hat{AUC_2}$ are the respective area estimators of the areas for the two cases, $SE(\hat{AUC_1})$ and $SE(\hat{AUC_2})$ are their respective standard errors, and
 r is a quantity representing the correlation introduced between the two areas from using the same sample training set.  The standard errors are given by

\begin{displaymath}
        SE(\hat{AUC})=\sqrt{\frac{\hat{AUC}(1-\hat{AUC})+(N_{GRS}-1)(Q_1-\hat{AUC}^2)+(N_{nonGRS}-1)(Q_2-\hat{AUC}^2)}{N_{GRS} \cdot N_{nonGRS}}}
\end{displaymath}
where estimators for $Q_1$ and $Q_2$ are $Q_1 \simeq \hat{AUC}/(2-\hat{AUC})$ and $ Q_2 \simeq 2 \cdot \hat{AUC}^2/(1+\hat{AUC})$ and $N_{GRS}$ and $N_{nonGRS}$ are the number of GRS and non-GRS in the training set, respectively.
A table in the reference then gives r as a function of the average of two areas  vs. the average of $r_{GRS}$ and $r_{non\-GRS}$,
where $r_{GRS}$ is the correlation coefficient between the estimated probabilities for the GRS in each area and $r_{non\-GRS}$ is the correlation coefficient between the non-GRS for each area.  They suggest Kendall's tau (see \citet{Press92} for an algorithm) as the appropriate correlation statistic for results obtained from an ordinal scale, as the case here.
To test the null hypothesis that the areas are the same for both cases, the ratio z is calculated as
\begin{eqnarray}
                                      z =  \frac{ \hat{AUC_1} - \hat{AUC_2} }{SE(\hat{AUC_1} -\hat{AUC_2})}
\end{eqnarray}
and compared to the value of $z_c$, the critical value for the probability of Type I error chosen.

\section{TRAINING SET RESULTS}
This section presents training set results for variations in pruning, skew, impurity measures, order of coefficient perturbation, number of hyperplanes per node examined, and number of jumps attempted to escape local minimum.  A feature set comparison is also made and GRS outliers are discussed.

\subsection{\it Pruning portion}
Initial attempts with training sets of larger skew showed that it was necessary to avoid pruning to obtain a classifier solution.  Subsequently, it was learned that \citet{Provost03}, in a study of decision tree induction for probability-based rankings, found larger trees may be better for probability estimation.  Thus, subsequent results have all been generated without pruning.  

\subsection{\it Training Set Skew-Ratio Comparisons}
While experience suggested OC1 is fairly robust, there was concern about its ability to construct decision trees with such high skew.
Thus it was decided to begin with examination of synthetic training sets with smaller number of non-GRS sources to explore the effect of skew on the quality of resulting decision trees.
Tests were run with modified training sets with 48, 1000, 5000, 17000, and 34000 non-target source pairs, as well as the full training set.  The AUC, and average probability estimate and maximum probability estimate of the training set GRS pairs were determined.  Results are shown in Table~\ref{skewtable}. 
  Calculation times for training set construction (with an ensemble of 50 decision trees) ranged from a few minutes for the smaller sets to a few hours for the full training set on an x86\_64 processor.

\begin{figure*} [p]
  \centering
  \plotone{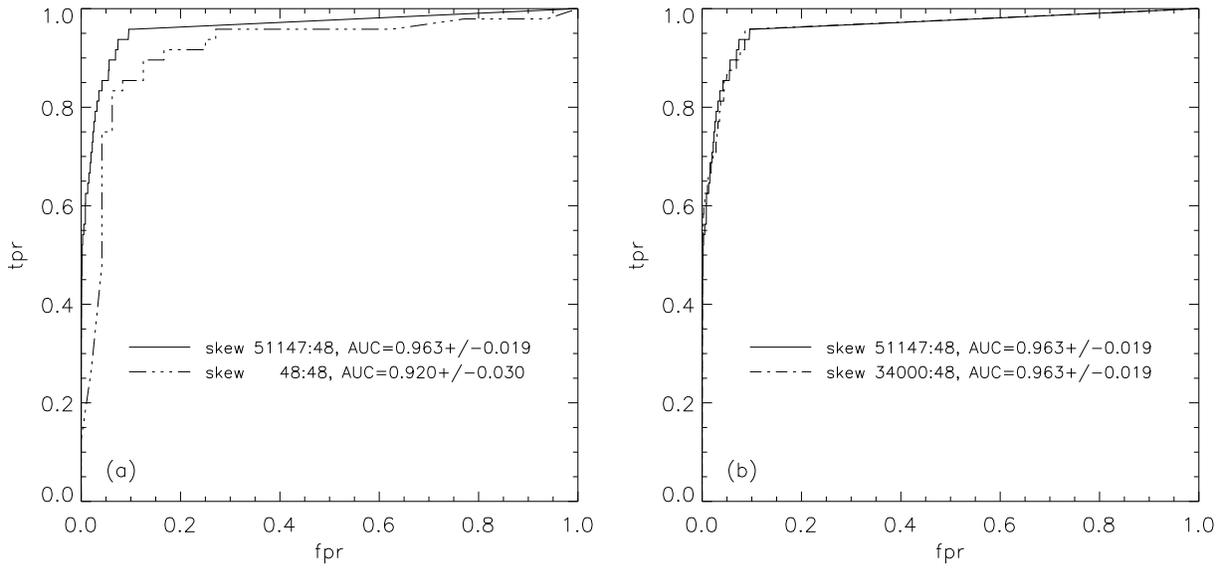}
  \caption{(a) A ROC curve comparison of training sets with skew 51147:48 and skew 48:48.
           (b) A ROC curve comparison of training sets with skew 51147:48 and skew 34000:48.} \label{roc_comp_skew}
\end{figure*}

Figure~\ref{roc_comp_skew}(a) shows ROC curve comparisons between the highest-skew and lowest-skew training sets and Figure~\ref{roc_comp_skew}(b) shows the comparison between the two highest skew sets examined.
For intermediate skew ratios, not shown, various amounts of interweaving of curves occured.  
However, it is noted that there appears to be a monotonic trend of improving AUC as skew ratios approach the inherent skew ratio for the population.  

Application of the \citet{Hanley83} formula, Eqn. \ref{sedif_auc} above, for the standard error of the difference in areas of two ROC curves, for the 48:48 skew curve and 51147:48 skew curve resulted in z=1.22.  Thus the hypothesis of no significant difference is accepted at the 5\% level.  However, the 48:48 skew curve being entirely below the 51147:48 curve, whereas the others interweave suggests it is more marginal.  A clear trend of degrading mean probability estimate and maximum probability estimate for the training set GRG as the skew ratio increases is apparent.
\cite{Provost03} in a study of tree induction for probability based rankings noted decision trees have been found to provide poor probability estimates, but discussed why decision tree representation is not intrinsically inadequate and recommended they be considered when rankings based on class membership is desired.  
\cite{Margineantu01} show probability estimation trees produce surprisingly good rankings, even when the probability estimates themselves are questionable.  
In Section 4 below, vote curve comparisons will be made that provide greater distinction between the various skew-ratio cases.

\subsection{\it Impurity Measure Comparisons}
\begin{figure*} [p]
  \centering
  \plotone{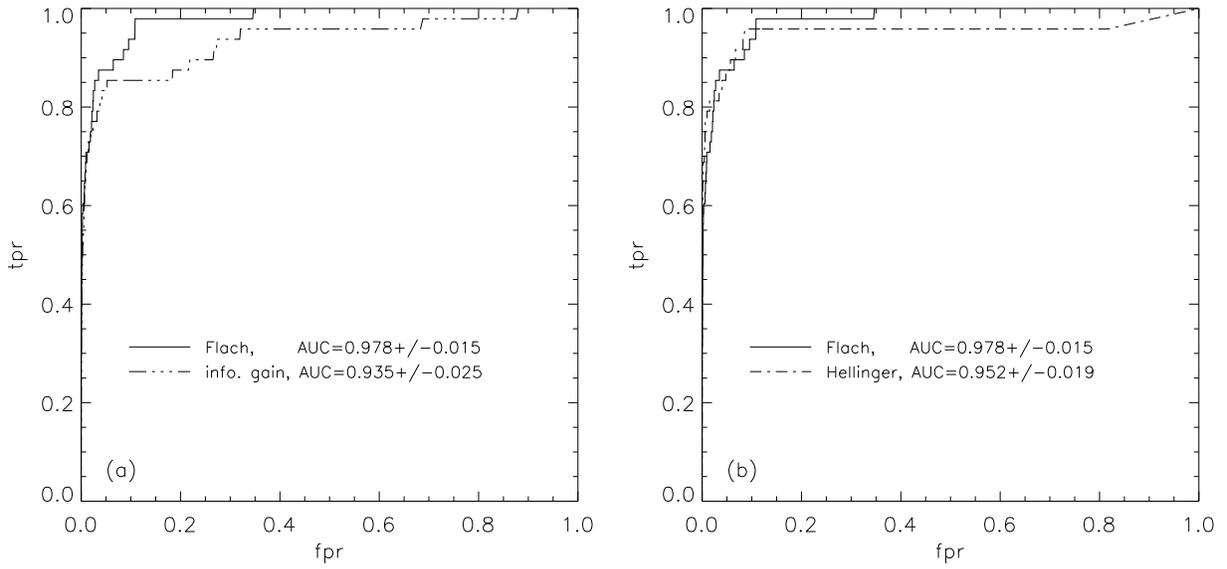}
  \caption{(a) ROC curve comparison of classifiers using Flach and Information-Gain impurity measures.  (b) ROC curve comparison of classifiers using Flach and Hellinger impurity measures.} \label{roc_comp_impurity}
\end{figure*}

Table~\ref{impuritytable} shows the results of utilization of the various impurity measures using the full training set.  
Figure~\ref{roc_comp_impurity}(a) shows ROC curve comparisons for the Flach and Information-Gain impurity measures, the two extreme AUC values for impurity measure tests and Figure~\ref{roc_comp_impurity}(b) shows the comparison between Flach and Hellinger impurity measures.  The remaining impurity measures compared with Flach also show considerable interweaving of the ROC curves, suggesting that while there is little to distinguish Flach from Gini, Variance, Twoing Rule or Hellinger, the difference between Flach and Information-Gain impurity measures may be more marginal.

Applying the formula for the standard error of the difference in areas of two ROC curves, for $\hat{AUC}(Flach)-\hat{AUC}(information\hspace{0.5em}gain)$, the highest and lowest AUC in the table, results in z=1.60. This is less than the critical $z_{c}=1.96$ for two sided value for normal distribution, for 5\% probability of Type I error.  Thus the hypothesis of no significant difference between these two ROC areas is accepted.  

\subsection{\it Order of Coefficient Perturbation Comparison} \label{Order of Coefficient Perturbation}
The order-of-coefficient-perturbation option determines the order in which the hyperplane coefficients are adjusted in searching for impurity measure improvement. 
Table~\ref{coefpertordertable} shows an order-of-coefficient-perturbation comparison.
A difference of AUC comparison of best-first and random order gave $z=1.40 < z_{c}$ indicating the hypothesis of no significant difference is accepted at the 5\% level.
It is noted however, -R12 option (loop 12 times picking a random coefficient and attempt to perturb it) also resulted in significantly longer computation times.

\subsection{\it Number of Hyperplanes per Node Comparison} \label{Hyperplanespernode}
 The number of restarts, i, is the number of hyperplanes searched for the best position at every node of the decision tree.  The best axis parallel hyperplane and (i-1) random hyperplanes are used. 
A difference of AUC calculation between using 20 and 60 hyperplanes/node resulted in accepting the hypothesis of no significant difference. Of course the more hyperplanes per node used the longer the decision tree generation run times.

\subsection{\it Number of Jumps attempted to escape local minimum}
Doubling the number of jumps, from 20 to 40 did not provide significantly different results.

\subsection{\it Feature Set Comparisons} \label{feature_set_comp}
An eight feature set was compared to the six feature set using the Flach impurity measure.  The two additional features used were the ratio of minor to major axes for each component.  There was no significant difference in AUC at the 5\% level.

\subsection{\it Outliers}
The sources J0317+769, J0342+636, J0508+609, J1036+677, J1847+707, J1951+706 showed votes consistently near zero over the various decision tree runs.  Further examination of these sources suggested J0317+769 and J0508+609 may have had possible wrong choices for training set pairs for the lobes and J1847+707 has possible contamination from a unresolved chance projections.  The poor results for J0342+636, J1036+677, and J1951+706 remain unclear, though contamination by chance projections remain a possible problem.  
Also, it is noted that while training set GRS have been lumped into one class, different types of GRS are expected to occupy different regions of feature space, a la FRI and FRII types. The small number of training set GRS may well be insufficient to distinguish these regions.


\section{APPLICATION SET RESULTS}

The candidate pairlist for the entire catalog was constructed with the same size constraints used for the training set and consisted of 870,370 pairs, of which 817,603 were below $+60\arcdeg$ declination.
This section discusses types of contamination found, makes a comparison of training set and application set vote curves and finally presents a table and mosaic of higher-ranked sources.

\subsection{\it Contamination}
Initial examination of images and contour plots of higher-ranked sources showed significant contamination from galactic plane sources, including high probability HII regions, about four dozen ring and shell type sources, most previously identified as Super Nova Remnant (SNR), and some apparent chance projections.  Also found were side-lobe artifacts from certain bright sources.

\begin{figure*} [p]
  \centering
   \plotone{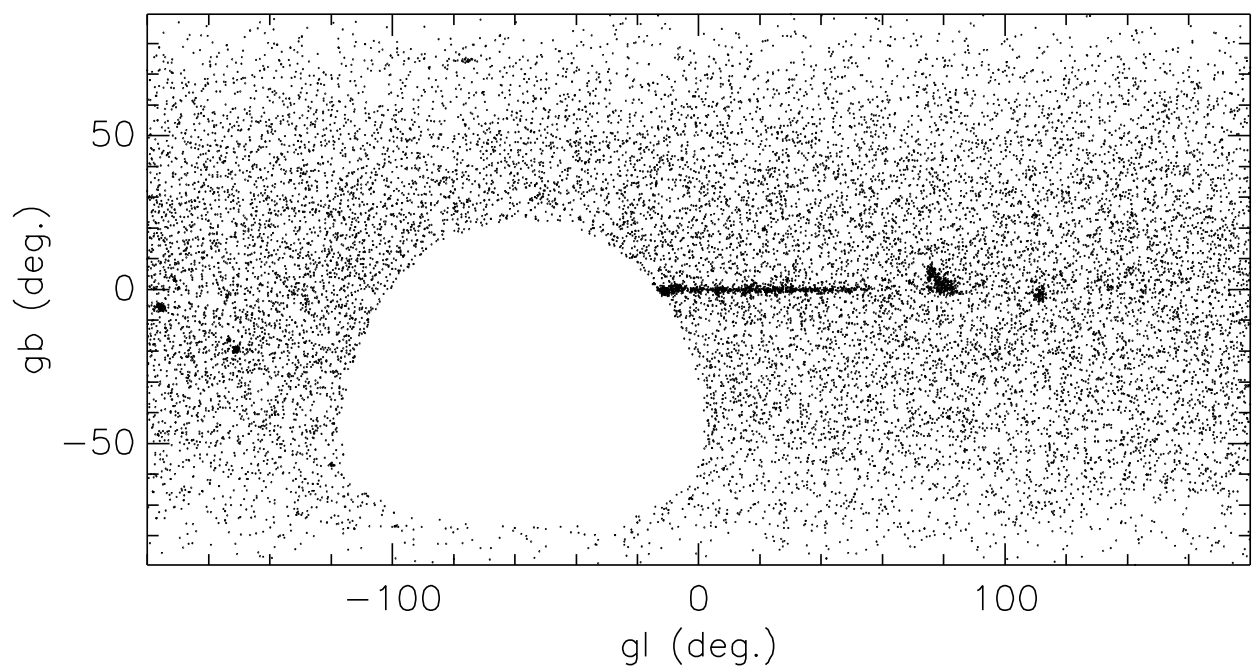}
  \caption{\label{fig:GalCoords_lgsilsz}Galactic coordinates of catalog sources with silhouette size $\geq$ 2.438 square arc minutes.  A random selection of 10\% of points are shown for improved plot clarity.} \label{GalCoords_lgsilsz}
\end{figure*}

A plot of the galactic coordinates of sources with silhouette sizes greater than 2.438 square arc minutes (the minimum size of at least one component for pairlist inclusion) for the  entire catalog are shown in Figure~\ref{GalCoords_lgsilsz}. 
It is noted that the ratio of large silhouette size sources to total number of sources for the training set and application set  differed by about five percent.  That difference may help account for excess galactic plane sources in the higher-ranked results.  Additionally, the current feature set may not be sufficient for discriminating against these galactic plane sources.

\begin{figure*} [p]
  \centering
  \plotone{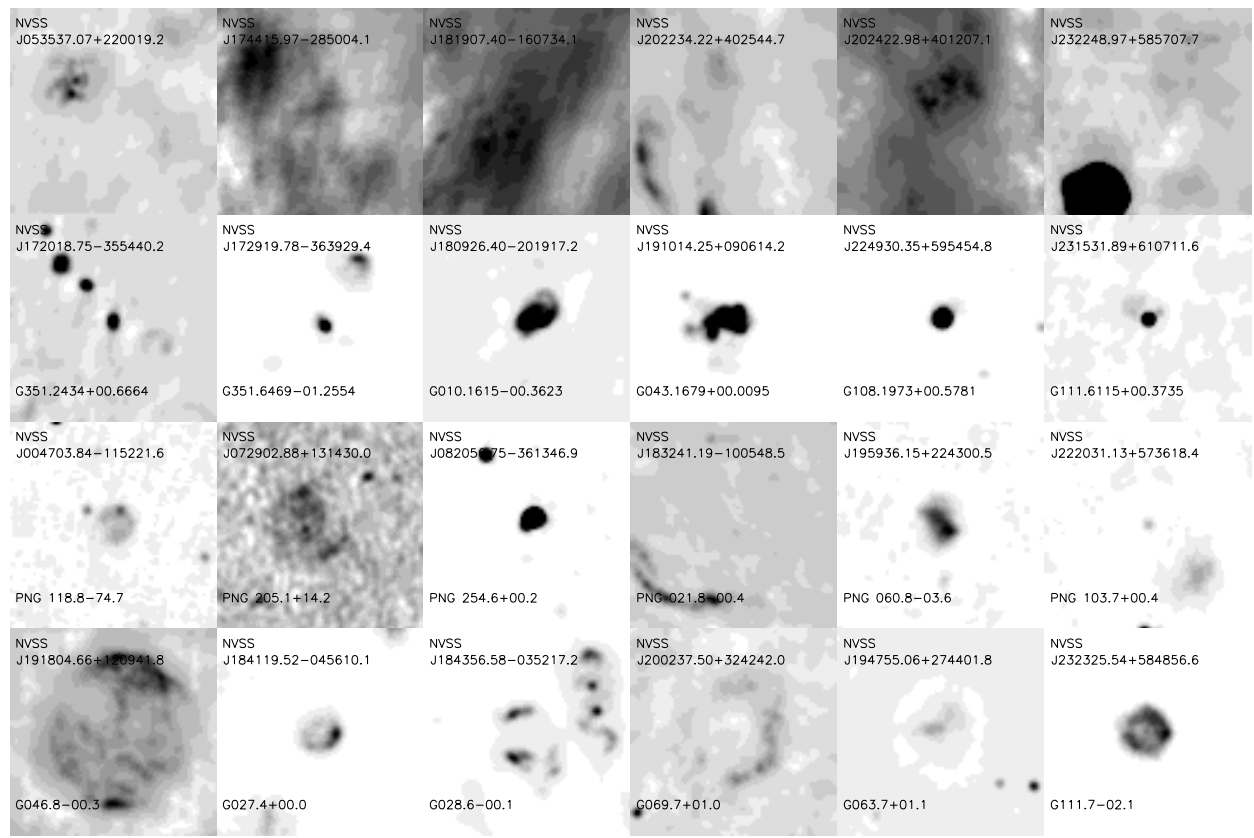}
  \caption{\label{fig:mosaic_contamination} Examples of 'contamination' and possible contamination in the training and application sets.  First row: Random selection of NVSS cutouts of galactic plane sources from higher-ranked pairlist.  Second row: Example NVSS cutouts of larger silhouette size \citet{Giveon05} catalog high reliability compact HII sources with NVSS coordinate and the MSX6C designation listed by \citet{Giveon05}.  Third row: Example NVSS cutouts of larger silhouette size \citet{Acker92} catalog high probability planetary nebulae with NVSS coordinate and PNG designation.  Fourth row:  NVSS cutouts of known SNR from higher-ranked pairlist with \citet{Green09} designations.  Image cutout size is $20\arcmin$ square.} \label{mosaic_contamination}
\end{figure*}

Figure~\ref{mosaic_contamination} shows example cutouts of some of this contamination.
A random selection of example higher-ranked source pairs from the galactic plane are shown in the first row of Figure~\ref{mosaic_contamination}.
The second row of Figure~\ref{mosaic_contamination} shows a selection of NVSS cutouts of larger angular size entries from the \citet{Giveon05} catalog of Galactic radio compact HII regions at 1.4 GHz. These were chosen to demonstrate the various morphologies of these sources in NVSS.  It is noted that it is possible for pairs of  HII regions to mimic GRS.  
NVSS images of 890 high probability planetary nebula from \cite{Acker92} were separately examined.  While perhaps 8\% showed no NVSS signature and most were point-like sources, a few had resolved morphology.  The third row of Figure~\ref{mosaic_contamination} shows a selection of NVSS images of larger angular diameter Acker sources for comparison.  We note that only one planetary nebula (NGC 1514) was identified in the higher-ranked source pair list.  
Finally the fourth row of the figure shows NVSS images of known SNR with \citet{Green09} designations recovered from the higher-ranked sources.
Of the ring and shell type sources, it is noted that 15 were subsequently identified as known SNR.  The remainder were an assortment identified in SIMBAD with HII regions, young stellar objects, dark nebulae, and  bubbles.
More features may be needed to discriminate against these sources.

\subsection{\it Vote Curve Comparison of the Training and Application Sets} 
\begin{figure*} [p]
 \centering
 \plotone{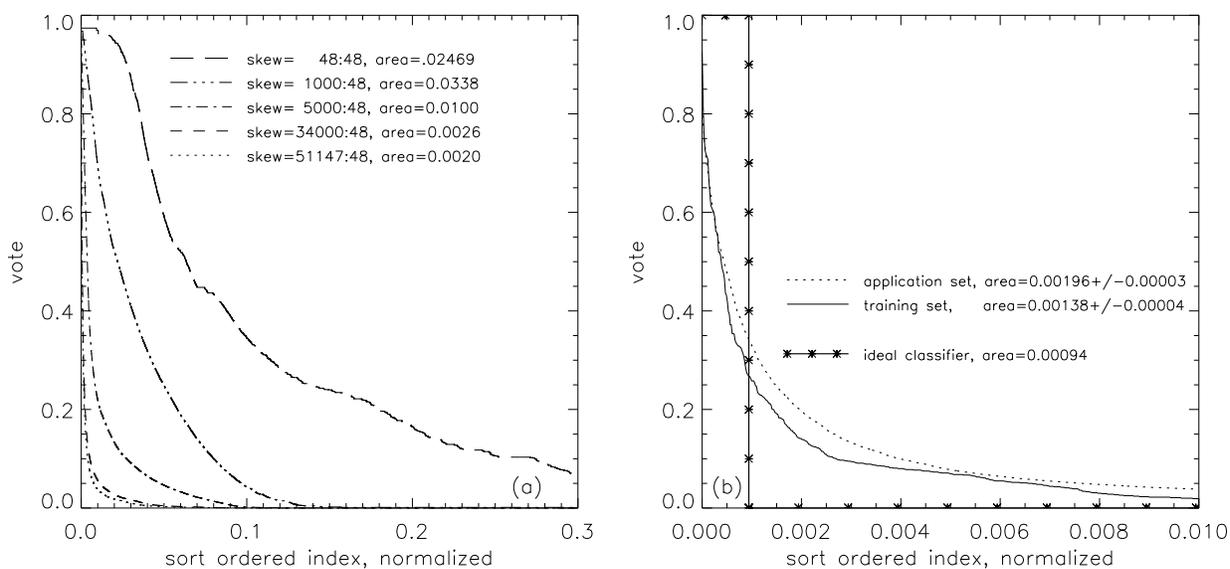}
 \caption{\label{fig:skew_app_votecurves}(a) Application Set Vote Curves from Training Sets with Various Skew.  Galactic plane sources and artifacts around certain bright sources have been excluded. (b) Vote curve comparison of skew 51147:48 training set and its corresponding application set.  Again galactic plane sources and artifacts around certain bright sources have been excluded from the application set.  The areas shown are for the total area under the respective vote curve.  The expected ideal vote curve for the population is also shown.} \label{skew_app_votecurves}
\end{figure*}

Initial evaluation of the area under the vote curve for all pairs below $+60\arcdeg$ declination resulted in a determination of $0.00227\pm0.00005$.  Excluding galactic plane sources and sources around certain bright sources resulted in an improved area determination of $0.00196\pm 0.0003$, an improvement of over five standard deviations.  Thus for further application set vote curve comparisons, pairs above $+60\arcdeg $ declination (the training set) and pairs within $5\arcdeg$ of the galactic plane were excluded.  Also excluded were pairs within $90\arcmin$ of Cyg A, M1, M42 and M87, areas with significant artifacts, leaving 438,350 pairs.  Vote curves for this restricted application set, using decision trees from the training sets with 48, 1000, 5000, 34000, and 51147 nonGRS, are shown in Figure~\ref{skew_app_votecurves}(a).  As the training set skew increases, they show a clear trend of improvement towards the target area under the curve of $48/51195 \simeq .00094$.  Note that for a perfect classifier, the vote curve would have a constant value of 1.0 at $x=0$, then drop to 0.0 at the normalized index of $\sim 0.00094$.  For greater detail, comparison of the 51147:48 skew restricted application-set vote curve with the 51147:48 skew training-set vote curve over the initial index range is shown in Figure~\ref{skew_app_votecurves}(b).  Figure~\ref{skew_app_votecurves} clearly demonstrates the need to compare vote curves for the training set and application set when the population's underlying skew ratio is unknown.  Of course, for sufficiently strong features, this skew effect may be obviated.  

A possible explanation for the difference between the the training set vote-curve area and the ideal-classifier area may be found in those resolved GRS sources that were excluded from the training set.  As will be discussed below, 19 were found to be higher-ranked sources when the decision trees were applied to their largest components.  If those are included, the the target area would become (48+19)/(51195+19)=.00131, a near match to the training set vote-curve area.

There are still some differences between the 51147:48 application set and training set vote-curve areas.  Possible reasons include false positives in the training set due to, for example, unresolved chance projections in the GRS class.  Also possible are unidentified true GRS labeled nonGRS.
There do not appear to be any GRS source pairs in the \citet{Lara01} sample for which both lobes are represented by approximate point-spread functions.  It may be that such pairs, though possible GRS, were not included in the original pre-selection.

\subsection{\it Higher-Ranked Candidates - Table and Mosaic}
Images and contour plots of over four thousand higher-ranked sources were examined.  These sources were produced using several of the better decision-tree ensembles.  For the pairs above $+60\arcdeg$, the rankings from the training-set decision trees were used.
Preliminary work was done to eliminate known HII regions, SNR, ambiguous galactic plane sources, and sources for which FIRST counterparts indicated chance projections.  
Table~\ref{grstable} presents the coordinates of 1616 GRS and candidate GRS from the higher-ranked sources, along with size and preliminary morphological annotations.  The key to the annotations are given in Table~\ref{masterkey}.
Typically, GRS/GRG are assigned an FR type.  For this work, FR classification was not attempted.  Though many of the larger 'd' and 't' annotated sources are clearly FRII, the smaller, less resolved sources were difficult to evaluate.  It was felt such distinctions deserved more uniform analytic attention.
In Table~\ref{grstable}, other morphological types were retained.  The annotations can be used to select promising candidates of morphologies of interest.
A few incidental identifications, not from a candidate pair list are also included, and annotated as such.

While well over a thousand of the sources in Table~\ref{grstable} appear to be viable GRS candidates, approximately one quarter appear to be resolved cores, chance projections, or ambiguous cases. They are included for possible future investigations.  Note a few of the systems in this table are less than $4\arcsec$, but also are included for documentation purposes.  Also, it should be noted that not all sources in this table have both source pair members as components. This particularly applies to the  resolved-core type morphology systems.

\begin{figure*} [p]
  \centering
  \plotone{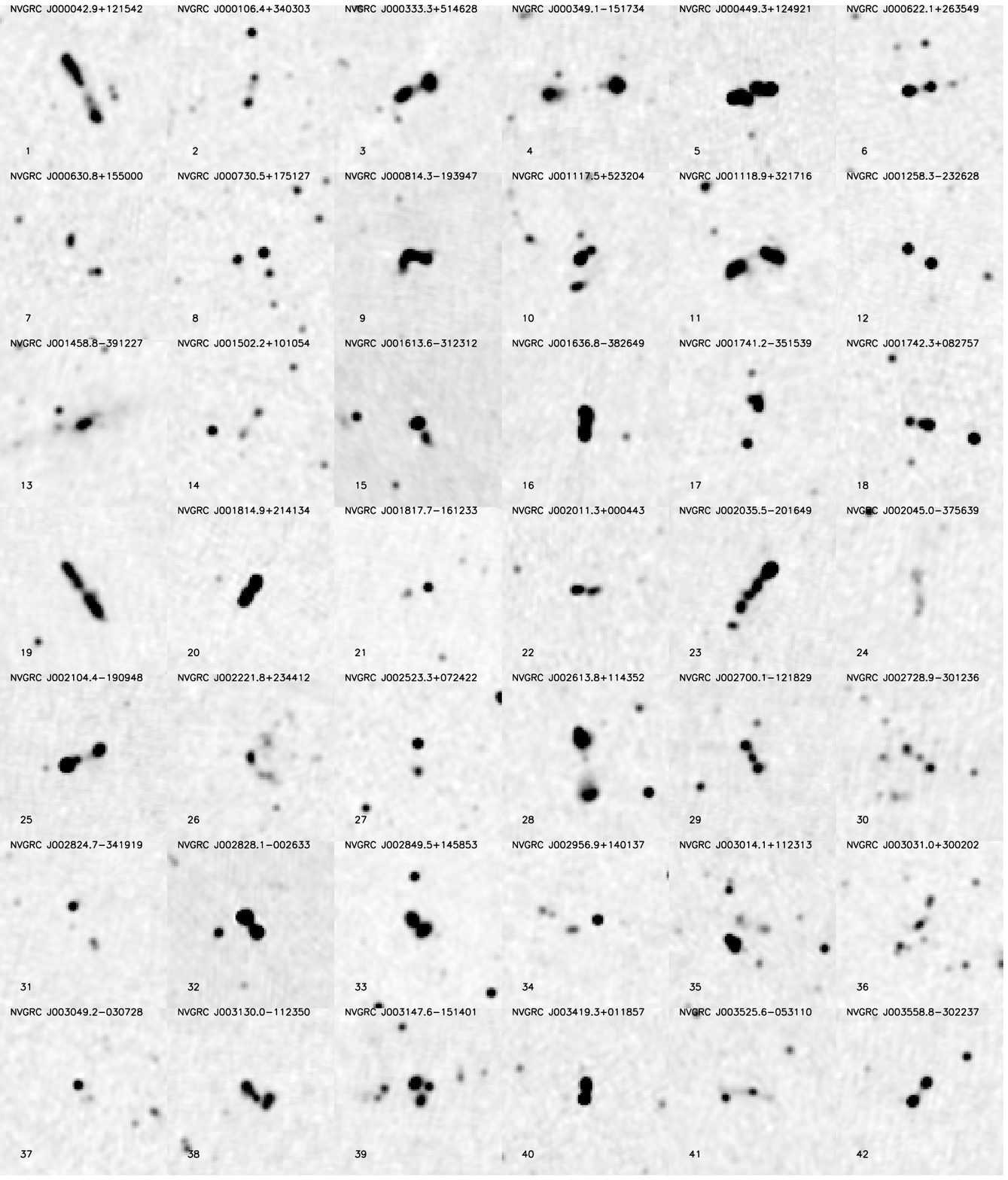}
  \caption{\label{fig:mosaic_table_GRSc} Giant Radio Sources and Candidates with IAU designation and Table 6 index.  These include some higher-ranked systems with ambiguous morphology as well as resolved core, core-jets and presumed chance projections, retained for documentation purposes.  Unless otherwise noted the images are $20 \arcmin$ square.  Previously identified GRS of \citet{Lara01}, \citet{Solovyov14}, and \citet{Machalski01} were not given another IAU designation.  The complete mosaic is available in the online version of the paper.} \label{mosaic_table_GRSc}
\end{figure*}

The significance of the system coordinate needs to be considered in the context of the system morphology and the estimated size, i.e. only the tabulated coordinates that are associated with an NVSS catalog source have the significant figures of the NVSS catalog itself.  The remainder are given to the same number of digits as the NVSS catalog for tabular consistency and convenience.
In this table, coordinate type 'c' indicates a NVSS catalog source presumed to be the core, and coordinate type 'v' is an estimate for the core position using lobe coordinates.  These latter, in general, were the average of the component positions, though some may have had visual adjustments.  
Figure~\ref{mosaic_table_GRSc} is a sample of those candidates, including those judged to have chance projection or otherwise ambiguous appearance, for which future observations could be made.  The complete mosaic is available in the online version of the journal.

While no comprehensive attempt was made at cross-identifications, those \citet{Lara01}, \citet{Solovyov14}, and \citet{Machalski01} GRS recovered were retained in the table, but not given another IAU designation. The coordinates listed for these systems are as constructed for the other candidates in the table. 
It is noted that of the 50 reported \citet{Solovyov14} GRS, 43 are in this candidate list.  (J185618.6+013120, though in their table of 51 entries, is a galactic plane source and was dropped from their further discussion.)  Of their remaining 7 GRS, four were resolved sources not expected to be found by this method, two (J172331.0-352542, J182708.3-124020) were perhaps more questionable galactic plane identifications, and one (J122045.0+055204) appeared to be a sidelobe of J121925.2+054945.9.
It is of interest that 19 of the 38 GRS of \citet{Lara01} sample that were excluded from the training set due to their resolved lobes, never-the-less, were recovered.  The overall GRS recovery rate from the Lara sample using a single ensemble is approximately half of such systems.  The recovery rate for the FRII and FRI/II systems of Machalski was only 40\%, perhaps reflecting the differing flux density limits for the Lara and Machalski samples.

\section{SUMMARY and CONCLUSIONS} 

Pattern recognition procedures have been developed for recovering candidates for a significant fraction of the expected number GRS from the NVSS catalog data.  An overview of the classification process was given.  
Training set construction, in particular the effect of skew was examined.  
ROC curve analysis and AUC comparisons, along with vote-curve comparisons were used to evaluate the classifiers.

The results of a selection of classifier options were compared.  It was found necessary to avoid pruning.  This is consistent with the observations of \citet{Provost03} that larger trees may provide better classifications. 
 The impurity measure comparisons showed that choice of impurity measure did not have a strong effect on classifier quality.

The need for vote curve comparison of the training set and application set when population skew ratios are unknown was demonstrated.  The best consistency between the training set and application set was for training-set skew ratios that more closely approached the inherent ratio.
The AUC and vote curve analysis suggested that for the given feature set, OC1 was quite adequate for the higher-skew training sets.

Cutouts of the higher-ranked source pairs were examined and preliminarily screened to eliminate known HII regions, SNR and chance projections, as well as ambiguous galactic plane sources.  A table of GRS and GRS candidates was constructed.  These include doubles and triples, and WAT, NAT, W, X, DD and S/Z-shaped systems.  Some GRS candidates with more PSF-like lobes were found in the candidate list, but since they were not represented in the training set, they are not expected to be well represented in the high probability sources.  

While classifiers have been generated that produce greater than 95\% probability that a randomly selected (GRS,non-GRS) pair will be properly ranked, probability estimates show considerable room for improvement and thus, are an area for future study.
Perhaps such low average probability estimates for training-set GRS are not surprising, given the relatively small number of training set GRS, the possibility of false positives and false negatives in the training set, and the grouping of GRS types into a single class.  However, a baseline for future work has been established.
  A more profitable line of inquiry may be to apply pattern recognition to separating the morphological types using the enlarged database and then revisiting the GRS vs. nonGRS classification.
A further area for study involves investigation of more powerful statistical tests for AUC comparison.  

With a larger database, there is promise of being able to sort types of radio galaxies and identify their regions in parameter space.  Future work involves exploring options for including source shape as well as core and redshift information and examination of the decision tree ensembles to determine what they reveal about the locations of various morphologies in feature space. 
While it is not impossible to visually search the entire set of NVSS maps for particular morphologies, it is believed  pattern recognition techniques will be needed to fully exploit information contained in the catalog and images.   Observationally verified training sets will be needed and the importance of a representative training set should be emphasized. 

**************************************************************************

\acknowledgments

The author is grateful to the anonymous referees for comments, suggestions, and a reference, which led to significant improvements of the paper.  Thanks are due to NRAO for page charge support.  The author appreciates office space and computing facilities provided by Lawrence Livermore National Laboratory's Institute of Geophysics and Planetary Physics (IGPP), John Bradley, director, during first phases of this work and Kem Cook, during later stages, as well as office space and computing facilities provided by B Division and host Tom McAbee, also of LLNL.  The author also appreciates further support provided by James Trebes, Physics Division, LLNL.  Thanks are also due Bill Cotton for software to access the NVSS images and Robert Becker for computer resources.

Extensive use was made of the IDL astro package (http://idlastro.gsfc.nasa.gov), GDL (http://gnudatalanguage.sourceforge.net), and IDL (a registered trademark of Exelis Visual Information Solutions).  The free availability of the OC1 decision tree software \newline (http://ccb.jhu.edu/software/oc1/oc1.tar.gz) was greatly appreciated.  The NVSS data was taken by the NRAO Very Large Array.  The National Radio Astronomy Observatory is a facility of the National Science Foundation operated under cooperative agreement by Associated Universities, Inc.  
This research has also made use of NASA's Astrophysics Data System (ADS), the NASA/IPAC Extragalactic Database (NED) which is operated by the Jet Propulsion Laboratory, California Institute of Technology, under contract with the National Aeronautics and Space Administration, and the Sloan Digital Sky Survey (SDSS), as well as the VizieR catalogue access tool, CDS, Strasbourg, France and the SIMBAD database, operated at CDS, Strasbourg, France.

This work was performed under the auspices of the U.S. Department of Energy by Lawrence Livermore National Laboratory in part under Contract W-7405-Eng-48 and in part under Contract DE-AC52-07NA27344.

\clearpage


\begin{appendix}
\section{APPENDIX MATERIAL}

Table~\ref{masterkey} provides the key for annotations in Table~\ref{grstable}, the table of GRS and candidates. 
It should be noted the annotations are not necessarily mutually exclusive, for example an S or Z-shaped source may also be a double-double candidate, thus groups may have multiple morphological classification types listed.  If multiple morphological types are given, they generally are given in order of estimated probabilities.

While most keyed items in Table~\ref{masterkey} are self-explanatory, a few comments are in order.

The 'cj' annotation refers to two components, one larger and brighter, the other smaller size and possible curved morphology.  The 'd' annotation consists of two components of similar silhouette size and peak flux.  The distinction between the two groups and becomes difficult for smaller and fainter systems.

The 'dd' annotation refers to a pair of double radio sources with a common center.  Some of these are candidate double double radio galaxies.

A few systems were noted for which the lobes appeared to have different FR morphologies, one diffuse with brightest region within the inner half of the radio source, the other with brightest regions near the outer edge.  See \citet{Gopal-Krishna2000} for detailed discussion of HYMORS.

The annotation 'irr' denotes a group whose members appear to be physically associated but are not well described by another morphology class.   Such groups may provide candidate merger or relic type sources. 

The annotation 'mg' (multiple possible groupings) was used to refer to groups for which multiple possibilities for physical associations appeared reasonable.  This annotation was followed by possible likely groupings.  These groups require higher resolution, sensitivity, and/or spectral data for definitive classification. 

The annotation 'rc' (resolved core) was used to describe a single contiguous non-pointlike, extended source that may or may not have associated jets or lobes apparent and may or may not have been resolved into multiple catalog components.  Their silhouette may be irregular.  Optical counterparts for some of these sources were found to be the centers of large galaxies.  


The distinction between double, resolved core, and core-jet morphology becomes rather arbitrary for smaller angular diameter groups.

The annotation 'unu' was generally applied to unusual configurations that could be explained by chance projection.  If they are not chance projections, they would be deemed interesting.

For the purposes of this paper, the annotation 'dl' (dogleg) is used to describe jets or lobe portions for which there is an apparent abrupt change in direction. 

The annotation 'int' (interesting) is a subjective evaluation, which may well have varied over the course of the classifications.

The comment 'tail' indicates a protrusion (normally low level) from lobe in direction away from core, whereas
 'jet' indicates a protrusion from a presumed core.

An 's' annotation for a source indicates the source appeared to be a source associated with the group by chance projection.  For some source components, the assignment of jet or lobe annotation was difficult and even arbitrary. 


\begin{deluxetable}{ll}
\tabletypesize{\scriptsize}
\tablenum{A-I}
\tablecaption{Key for Table 6 \label{masterkey}}
\tablewidth{0pt}
\tablehead{ \colhead{Code} & \colhead{Classification or Description} }
\startdata
 amb&   ambiguous\\
 arc& arc or C-shape, possible edge brightened lobe, bent jets, or ring fragment\\
 cj&  core-jet\\
 d&   double lobe radio galaxy - may include core-jet sources\\
 dd&  double double morphology (DD), may be candidate DDRG\\
 dl&  dogleg\\
 fishtail&  multiple (usually two) lower level tails\\
 hymor& hybrid morphology radio source (HYMOR)\\
 int& interesting\\
 irr& irregular, distorted\\
 incidental& non-source pair systems found during cutout examinations\\
 jet& protrusion away from presumed core\\
 lobe& lobe, resolved lobe\\
 mg&  multiple groupings possible\\
 nat& narrow angle tail (NAT)\\
 point& non-extended source for which the major and minor axes appear to be the same size\\
 rc&  resolved compact source, non-pointlike\\
 ring& ring\\
 ring-lobe& edge or rim brightened lobe or embedded ring\\
 s&   single source (sometimes with sidelobes) or probable chance projection of point-like source into group\\
 sz&  S or Z-shaped\\
 t&   triple, no bend, little extended structure\\
 tail&  protrusion(normally lower level) from lobe in direction away from core\\
 unu& unusual, uncommon - may be due to chance projection\\
 w&   W-shape, wiggles\\
 wat& wide angle tail (WAT)\\
 x&   X-shape\\


\enddata
\end{deluxetable}
\clearpage

\end{appendix}

\clearpage

\clearpage
\newpage
\begin{deluxetable}{ll}
\tabletypesize{\scriptsize}
\tablecaption{List of Features for GRS Pattern Recognition \label{GRS_features}}
\tablehead{
\colhead{Feature} & \colhead{Description}}
\startdata
 $silsz_{l}$................ & silhouette size, component with larger silhouette size \\
 $silsz_{s}$................ & silhouette size, component with smaller silhouette size \\
 $flx_{l}$.................  & peak flux, component with larger silhouette size \\
 $flx_{s}$.................  & peak flux, component with smaller silhouette size \\
 $d$.......................  & source pair separation \\
 $N_{neighbors}$...........  & Number of sources within 20' of larger silhouette-size source having \\
 $             $             &   silhouette size larger than 1.871 square arc minutes \\
\enddata
\tablecomments{The silhouette size is the area enclosed by the fitted model at the 1.4 mJy level.}
\end{deluxetable}

\begin{deluxetable}{ccc}
\tabletypesize{\scriptsize}
\tablecaption{Contingency Table - Two Class Problem \label{conf_matrix}}
\tablewidth{0pt}
\tablehead{
\colhead{Positive Prediction} & \colhead{Negative Prediction} & \colhead{Totals}
}
\startdata
\tableline
 True Positive (TP) & False Negative (FN) & Positive Class (POS) \\
False Positive (FP) & True Negative (TN)  & Negative Class (NEG) \\
\tableline
  Predicted Positive (PP) & Predicted Negative (PN) & $N_{ts}$   \\
\enddata
\end{deluxetable}


\begin{deluxetable}{rccc}
\tabletypesize{\scriptsize}
\tablecaption{Synthetic Training Sets Compared with Complete Training Set \label{skewtable}}
\tablewidth{0pt}
\tablehead{
\colhead{Training Set Skew} & \colhead{AUC\tablenotemark{a}} & \colhead{mean\tablenotemark{b}} & \colhead{maximum\tablenotemark{c}} 
}
\startdata
48:48                   & 0.920 \underline{+} 0.030 &   0.862 \underline{+} 0.038 &   0.9760 \underline{+} 0.0002 \\  
1000:48                 & 0.920 \underline{+} 0.027 &   0.682 \underline{+} 0.055 &   0.9688 \underline{+} 0.0004 \\  
5000:48                 & 0.928 \underline{+} 0.026 &   0.474 \underline{+} 0.058 &   0.9601 \underline{+} 0.0006 \\  
17000:48                & 0.957 \underline{+} 0.020 &   0.330 \underline{+} 0.053 &   0.9588 \underline{+} 0.0010 \\  
34000:48                & 0.963 \underline{+} 0.019 &   0.296 \underline{+} 0.045 &   0.9484 \underline{+} 0.0007 \\  
51147:48                & 0.963 \underline{+} 0.019 &   0.230 \underline{+} 0.038 & 0.9269 \underline{+} 0.0057 \\  
\enddata
\tablecomments{Options: best first order of coefficient perturbation, 40 hyperplanes searched per node, no pruning, variance impurity measure}
\tablenotetext{a}{Area under ROC curve with standard error}
\tablenotetext{b}{Unweighted mean and standard deviation of mean for training set GRS}
\tablenotetext{c}{Mean of maximum vote with standard deviation of mean}
\end{deluxetable}



\begin{deluxetable}{lccc}
\tabletypesize{\scriptsize}
\tablecaption{Impurity Measure Comparisons \label{impuritytable}}
\tablewidth{0pt}
\tablehead{
\colhead{impurity measure} & \colhead{AUC\tablenotemark{a}} & \colhead{mean\tablenotemark{b}} & \colhead{maximum\tablenotemark{c}}
}
\startdata
\tableline           
Flach                   & 0.978 \underline{+} 0.015 & 0.191 \underline{+} 0.032 & 0.786 \underline{+} 0.083  \\  
Gini Index              & 0.965 \underline{+} 0.019 & 0.203 \underline{+} 0.032 & 0.898 \underline{+} 0.020  \\  
Variance                & 0.963 \underline{+} 0.019 & 0.230 \underline{+} 0.038 & 0.927 \underline{+} 0.006  \\  
Twoing Rule             & 0.962 \underline{+} 0.019 & 0.228 \underline{+} 0.042 & 0.929 \underline{+} 0.008  \\  
Hellinger               & 0.952 \underline{+} 0.021 & 0.313 \underline{+} 0.045 & 0.908 \underline{+} 0.021  \\  
Information Gain        & 0.935 \underline{+} 0.025 & 0.132 \underline{+} 0.022 & 0.563 \underline{+} 0.144  \\  
\enddata
\tablecomments{Options: best first order of coefficient perturbation, 40 hyperplanes searched per node, no pruning}
\tablenotetext{a}{Area under ROC curve with standard error}
\tablenotetext{b}{Unweighted mean and standard deviation of mean for training set GRS}
\tablenotetext{c}{Mean of maximum vote with standard deviation of mean}
\end{deluxetable}

\begin{deluxetable}{lccc}
\tabletypesize{\scriptsize}
\tablecaption{Order of Coefficient Perturbation Comparisons \label{coefpertordertable}}
\tablewidth{0pt}
\tablehead{
\colhead{Order of Coefficient Perturbations} & \colhead{AUC\tablenotemark{a}} & \colhead{mean\tablenotemark{b}} & \colhead{maximum\tablenotemark{c}} 
} 
\startdata
best first                      & 0.978 \underline{+} 0.015 & 0.191 \underline{+} 0.032 & 0.786 \underline{+} 0.083 \\ 
sequential                      & 0.966 \underline{+} 0.018 & 0.140 \underline{+} 0.025 & 0.602 \underline{+} 0.108 \\ 
random (-R12)                   & 0.942 \underline{+} 0.024 & 0.161 \underline{+} 0.026 & 0.606 \underline{+} 0.109 \\ 
\enddata
\tablecomments{Options: 40 hyperplanes searched per node, no pruning, Flach impurity measure}
\tablenotetext{a}{Area under ROC curve with standard error}
\tablenotetext{b}{Unweighted mean and standard deviation of mean for training set GRS}
\tablenotetext{c}{Mean of maximum vote with standard deviation of mean}
\end{deluxetable}

\notetoeditor{I couldn't figure out how to get fixed width fonts for this table so that the designations and coordinates in the tables would all come out the same width instead of ragged alignment.}
\notetoeditor{Machine Readable Table requested for Table 6 - tab6.txt. Stub follows.}

\begin{deluxetable}{ccccccl}
\tabletypesize{\scriptsize}
\tablecaption{Giant Radio Sources and Candidates \label{grstable}}
\tablewidth{0pt}
\tablehead{
\colhead{Index\tablenotemark{a}} & \colhead{\parbox{4em}{Source\\Designation}} & \colhead{R.A.(J2000)} & \colhead{Dec.(J2000)} & \colhead{\parbox{4em}{coordinate\\ type\tablenotemark{b}}} & \colhead{size (')} & \colhead{Annotations\tablenotemark{c}}
}
\startdata
   1 &  NVGRC J000042.9+121542 & 00 00 42.93 & +12 15 42.4 & v &  10.0 &    d  dd? \\
   2 &  NVGRC J000106.4+340303 & 00 01 06.40 & +34 03 03.0 & c &   5.0 &    t? \\
   3 &  NVGRC J000333.3+514628 & 00 03 33.34 & +51 46 28.9 & c &   7.0 &    t? asym core not apparent \\
   4 &  NVGRC J000349.1-151734 & 00 03 49.19 & -15 17 34.6 & c &  11.0 &    t \\
   5 &  NVGRC J000449.3+124921 & 00 04 49.31 & +12 49 21.5 & v &   7.0 &    mg d? 4' d with adj 4' d or irr? \\
\enddata

\tablecomments{Table~\ref{grstable} is published in its entirety in the electronic edition of this journal.  A portion is shown here for guidance regarding its form and content.}
\tablecomments{Approximately two dozen candidates less that $4.0\arcmin$ were retained in the table for documentation purposes.}
\tablecomments{Coordinate based source designations are truncated, not rounded.  Format=JHHMMSS.s+DDMMSS}
\tablenotetext{a}{For easy reference, the index corresponds to numbered cutouts in Figure~\ref{mosaic_table_GRSc}}.
\tablenotetext{b}{Coordinate type c indicates system coordinate is a NVSS fitted catalog entry, v indicates visual estimate from symmetry considerations.  It should be noted that only the coordinates associated with a NVSS catalog entry have the significance of the NVSS entry.}
\tablenotetext{c}{The key to the annotations is given in Table~\ref{masterkey}. Although the training set consisted of primarily FRI and FRII type sources, FR classification was not attempted.  Though many of the larger 'd' and 't' annotated systems are clearly FRII, the smaller, less resolved sources were difficult to evaluate.  It was felt such distinctions deserved more uniform analytic attention.  Also note that other morphological types were retained for documention purposes.  GRS previously identified by Lara et al. (2001) [2001A\&A...370..409L], Machalski et al. (2001) [2001A\&A...371..445M], and Solovyov \& Verkhodanov (2014) [2014AstBu..69..141S], were retained but not given an NVGRC designation.  Aside from the GRSs/GRGs of these three references and SNR identified by Green (2014,2016), no comprehensive attempt was made at cross-identifications.  Coordinates are given as for other entries.
}.
\end{deluxetable}%


\end{document}